# *ODIN*: A Bit-Parallel Stochastic Arithmetic Based Accelerator for In-Situ Neural Network Processing in Phase Change RAM

Supreeth Mysore Shivanandamurthy, Ishan. G. Thakkar, Sayed Ahmad Salehi

*Department of Electrical and Computer Engineering, University of Kentucky, Lexington, KY,USA.*
*supreethms@uky.edu*, *igthakkar@uky.edu*, *sayedsalehi@uky.edu*.

*Abstract*—Due to the very rapidly growing use of Artificial Neural Networks (ANNs) in real-world applications related to machine learning and Artificial Intelligence (AI), several hardware accelerator designs for ANNs have been proposed recently. In this paper, we present a novel processing-in-memory (PIM) engine called *ODIN* that employs hybrid binary-stochastic bit-parallel arithmetic inside phase change RAM (PCRAM) to enable a low-overhead in-situ acceleration of all essential ANN functions such as multiply-accumulate (MAC), nonlinear activation, and pooling. We mapped four ANN benchmark applications on *ODIN* to compare its performance with a conventional processor-centric design and a crossbar-based in-situ ANN accelerator from prior work. The results of our analysis for the considered ANN topologies indicate that our *ODIN* accelerator can be at least 5.8× faster and 23.2× more energy-efficient, and up to 90.8× faster and 1554× more energy-efficient, compared to the crossbar-based in-situ ANN accelerator from prior work.

## I. INTRODUCTION

Artificial Neural Networks (ANNs) have achieved remarkable progress in recent years, and they are being aggressively utilized in real-world applications related to artificial intelligence (AI) and machine learning [1]. In general, ANNs mimic biological neural networks, and utilize compute-heavy arithmetic functions such as multiply-accumulate (MAC), nonlinear activation, and pooling. Although these ANN functions are amenable to acceleration because of a high degree of compute parallelism, their acceleration is challenging because of the need to avoid the memory wall while accessing their large number of operands [4]. To address this problem, several prior works have explored crossbar memory-based processing-in-memory (PIM) accelerators (e.g., [6]-[11]) that leverage the Kirchoff's Law to perform MAC operations in the analog domain. However, such analog computing-based accelerators require power-hungry and sluggish digital-to-analog converters and analog-to-digital converters (DACs and ADCs), which diminishes the performance and energy-efficiency benefits of such accelerators. Moreover, these accelerators do not fully capitalize on the PIM paradigm, as they still have to heavily rely on conventional processor-centric computing for implementing essential ANN functions such as nonlinear activation and pooling. This motivates the need for simple, low-overhead, and energy-efficient in-situ accelerators that can realize the full potential of PIM-based ANN processing.

In this paper, we present a phase change RAM (PCRAM) based in-memory ANN accelerator called *ODIN*. *ODIN* uses stochastic arithmetic to convert complex MAC operations into a series of simple and low-overhead in-situ logical operations that are implemented using the analog computing capabilities of PCRAM [3]. Stochastic arithmetic typically requires additional circuits to enable conversion of the operands between the stochastic and binary number formats [5]. *ODIN* employs lightweight CMOS add-on logic inside PCRAM banks to implement such number conversion circuits with minimal area and power consumption overheads. *ODIN* also employs custom CMOS logic blocks to realize binary arithmetic-based implementation of nonlinear activation and pooling functions of ANNs. Stochastic arithmetic for ANN processing has been utilized before in the DRAM-based in-situ accelerator described in [5]. The accelerator from [5] employs heavy add-on logic inside DRAM banks, which can increase the total area of DRAM chips by up to 4× [21]. In contrast, *ODIN* presents the first of a kind stochastic arithmetic based accelerator that includes extremely low-overhead add-on logic, along with very lightweight modifications in PCRAM banks and PCRAM controller, for efficient processing of ANNs. Our key contributions in this work are summarized below.

- We present a novel processing-in-memory (PIM) accelerator called *ODIN* that leverages hybrid binary-stochastic arithmetic to accelerate ANN processing in PCRAM;
- *ODIN* employs low-overhead add-on logic and lightweight PCRAM modifications to implement stochastic arithmetic based MAC operations and binary arithmetic based activation and pooling operations directly inside PCRAM banks;
- We evaluate our *ODIN* accelerator for four ANN benchmark topologies (i.e., CNN1, CNN2, VGG1, and VGG2) from MLBench library[20], with two datasets MNIST and ImageNet;
- We compare the performance of our *ODIN* accelerator for the considered ANN benchmarks with the following architectures: baseline CPU-only with 32-bit floating precision, CPU-only with 8-bit fixed precision, as well as the pipelined and unpipelined variants of the ISAAC accelerator from [2].

## II. RELATED WORK AND MOTIVATION

Several processing-in-memory (PIM) based ANN accelerators are proposed in prior work (e.g., [2][5][12]-[20]). Some of these accelerators use memory in the conventional volatile RAM configuration (e.g., SRAM [15], DRAM [5] [12]-[14] based ANN accelerators), whereas some others use the emerging non-volatile memory in the crossbar configuration (e.g., STT-MRAM [18]-[19], ReRAM [2][16][17][20]). The RAM-configured ANN accelerators use bit-line computation to perform basic bit-parallel logical operations in a single read of a traditional RAM. Each memory column (bit-line) is further transformed into a bit-serial ALU by adding extra logic, multiplexing, and state elements in the peripheral circuitry, to perform complex arithmetic operations such as multiplication and addition, which are very essential for ANN processing [6][7]. However, these accelerators incur high area-power overheads in their augmented peripherals, and suffer from low throughout of bit-serial execution. Along the same line, a recently proposed DRAM based in-situ accelerator [5] uses stochastic arithmetic to perform ANN operations. However, this accelerator heavily modifies peripherals to facilitate logic circuitry, registers, shifters to enable carry-save addition required for MAC processing. These modifications can result in up to 4x area increase per DRAM bank [21]. On the other hand, NVM crossbar-based accelerators also suffer from high overheads of ADC and DAC circuits. In contrast, our proposed design *ODIN* stands first of its kind to leverage the hybrid binary-stochastic arithmetic inside PCRAM banks for implementing a very low-overhead acceleration of ANNs.

## III. BACKGROUND AND FUNDAMENTALS

### A. Artificial Neural Networks (ANNs)

To explain the structure and operations involved in an ANN, we explain a highly popular ANN model called Convolutional Neural Network (CNN). Generally, CNNs consist of majorly four types of layers (Fig. 1), namely convolution (CONV), local response normalization (LRN), pooling (POOL), and fully connected (FC) layers [7]. Among all these layers, the major portion of information of the input data gathering is taken place in the CONV and FC layers are computationally heavy, wherein a very large number of multiply-accumulate (MAC) operations occur between the synaptic weights and input activation values. In addition to MAC operations, other important operations in a CNN (ANN) are nonlinear activation and pooling. These ANN operations typically involve many steps for intermediate value preparation and storage, and therefore, they can be highly time and energy consuming. For better insight, consider Eq. (1), related to an FC layer, which consists of MAC operations between the weights ($w_{i,j}$'s) and input activation values ($a_i$'s). In Eq. (1), $w_{i,j}$ is the synaptic weight connecting $i^{th}$ node in layer '$a$' to $j^{th}$ node in layer '$b$', and '$f()$' is a non-linear activation function (e.g., ReLU [7]).

$$b_j = f(\sum_{V_i}(w_{i,j} * a_i))  \quad (1)$$

Moreover, popular pooling functions for CNNs are max and average pooling functions. Also, note that, since *ODIN* uses stochastic arithmetic, the results are always between '0' and '1'. Therefore, for processing CNNs (ANNs) on *ODIN* accelerator, *the LRN layer can be safely ignored with minimal loss of accuracy [2]*.

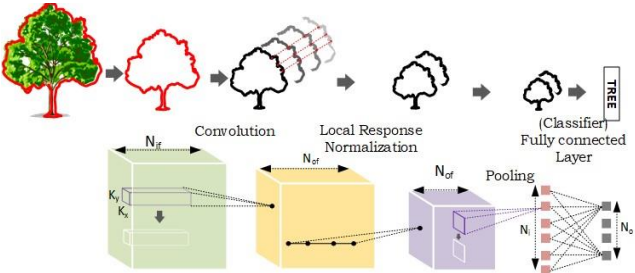

**Fig. 1: Four major types of layers found in Convolutional Neural Networks (CNNs) [7].**

### B. Phase Change RAM (PCRAM) Architecture

This section provides a brief background on PCRAM organization required to understand *ODIN*. The operation of PCRAM is not discussed here, for which the reader is encouraged to refer to prior work [1] and [22]. PCRAM, like DRAM, is organized hierarchically [22]. An example PCRAM memory of 16GB capacity has 2 channels, with 8 ranks per channel, and 16 banks per rank. A PCRAM bank has 16 partitions, each of which is an array of 4096 wordlines and 8K bitlines. A PCRAM bank has 256 peripheral structures, which include the sense amplifiers (S/As) to read and the write drivers (W/Ds) to write. The peripheral structures in PCRAM banks allow to read and program 256 PCRAM cells in parallel. Therefore, the read and write granularity is 256 bits (the size of a memory line).

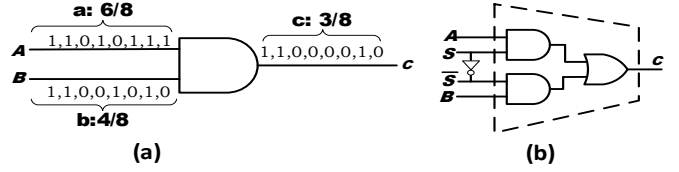

**Fig. 2: Stochastic arithmetic circuits. (a) Multiplier (AND gate), (b) Adder (MUX circuit).**

### C. Stochastic Arithmetic

In stochastic arithmetic, data is processed in the stochastic number (SN) format instead of the conventional binary number (BN) format. In SN format, information data is represented as pseudorandom bit streams. Using stochastic arithmetic, the complexity of performing MAC operations can be reduced substantially. For example, multiplication can be implemented by a simple bit-parallel AND operation, and scaled addition (accumulate) can be realized by a bit-parallel multiplexing (MUX) operation [23]. Fig. 2(a) depicts an AND gate implementing c = a×b in the SN format and Fig. 2(b) shows a MUX implementing c= s×a+s'×b in the SN format. In *ODIN* engine, we always use s=0.5 for MUX-based stochastic addition.

## IV. *ODIN* FRAMEWORK: OVERVIEW

The design of *ODIN* accelerator includes the following three hardware architecture attributes: (i) integration with heterogeneous computing system, (ii) lightweight modification of peripherals and inclusion of add-on logic in PCRAM banks to enable in-situ processing of ANNs, and (iii) introduction of new memory controller commands to support the processing-in-memory (PIM) functionality in PCRAM. Each of these attributes is described below.

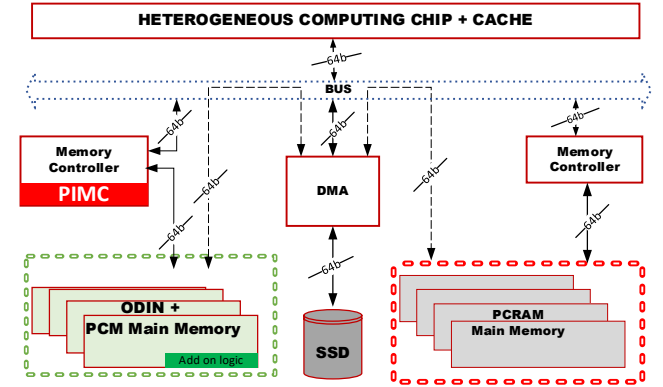

**Fig. 3: Integration of *ODIN* in a heterogeneous system.**

### A. Integration with Heterogeneous Computing System

Fig. 3 shows the heterogeneous computing system architecture that employs our *ODIN* accelerator. The system consists of a heterogeneous processing chip (HPC) that has heterogeneous cores (e.g., CPU, GPU cores) integrated with on-chip memory. The HPC connects, through a bus, with a phase change RAM (PCRAM) based main memory subsystem and a storage subsystem (SSD based). The PCRAM main memory subsystem has two channels. One of the PCRAM channels functions as the main memory channel by default. In contrast, the PCRAM modules on the other channel are lightly modified with add-on logic to constitute our *ODIN* accelerator, which has PIM capabilities for efficient in-situ acceleration of ANNs. These modified PCRAM modules, referred to as *ODIN* accelerator modules, also serve as PCRAM-based main memory modules when their PIM functionality is not in use. Both the *ODIN* accelerator and PCRAM main memory channels are connected to the storage channel through direct memory access (DMA) controllers, in addition to being connected to conventional memory controllers that manage regular memory operations. It is assumed that the DMA controller, in association with the software stack, can prefetch all the operands related to an ANN under execution from the SSD channel and load them in the *ODIN* accelerator modules, to subsequently trigger their processing directly inside the *ODIN* accelerator modules. The regular memory controller associated with the *ODIN* accelerator channel is also lightly modified with PIM controller (PIMC) capabilities, to support additional commands for efficient management of *ODIN's* operation.

### B. Hardware Modifications in PCRAM Banks

Our *ODIN* accelerator can fully process ANNs directly inside PCRAM. For that, *ODIN* employs lightweight modifications in PCRAM banks that enable in-situ execution of essential arithmetic functions of ANNs, such as multiply-accumulate (MAC), activation, and pooling. To reduce the overheads of effected modifications, *ODIN* employs hybrid binary-stochastic arithmetic for implementing these functions. To this effect, *ODIN* implements MAC operations using stochastic arithmetic, whereas for activation and pooling functions *ODIN* uses the traditional binary arithmetic. The specific modifications in the PCRAM bank structure, effected to implement various ANN functions, are described in the subsections below. In addition, *ODIN* also dedicates an entire partition per PCRAM bank as a scrap memory space that is utilized to help perform various ANN functions in situ. This partition is identified as Compute Partition in Fig. 4(a), which shows a schematic of *ODIN*'s PCRAM bank with effected modifications.

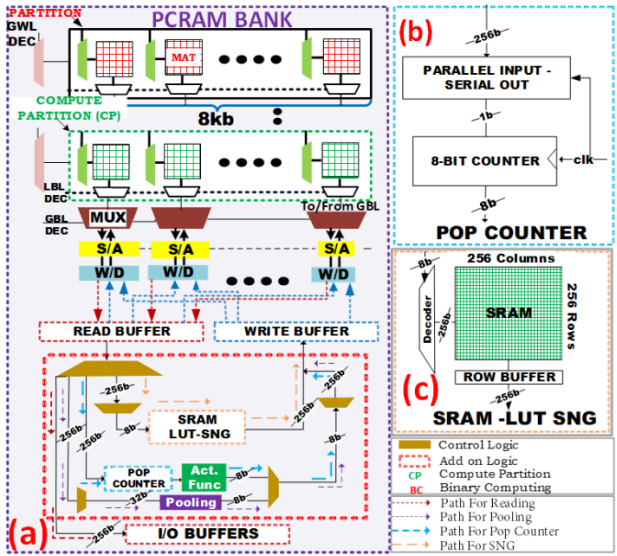

**Fig. 4: Schematic of a PCRAM bank modifications in *ODIN*.**

*1) Hardware for Multiply-Accumulate (MAC) Operations*

To implement in-situ MAC operations inside PCRAM, *ODIN* uses stochastic arithmetic, which transforms complex multiplication and addition operations, respectively, into simple bit-parallel AND and multiplexing operations. Multiplexing operations can be further divided into bit-parallel AND and OR operations (Fig. 2(b)) using stochastic arithmetic. Thus, *ODIN* employs stochastic arithmetic to convert complex MAC operations into a series of simple bit-parallel AND and OR operations. However, for MAC operations to take place inside PCRAM in the stochastic format, all MAC operands need to be present in PCRAM in the stochastic format. Storing all MAC operands of an ANN inside PCRAM ahead of time in the stochastic format may require impractically high storage capacity. This is because an N-bit binary operand typically occupies $2^N$-bits, when it is converted into the stochastic format with a reasonable precision. Therefore, to reduce the storage footprint of ANN MAC operands, *ODIN* makes two adjustments. First, it makes ANN MAC operands available in PCRAM banks in the binary format. Second, it fixes the operands size to 8-bits - this adjustment resonates with other prior works on PIM-based ANN accelerators (e.g., [7], [20]) wherein the size of ANN operands is fixed without significant loss of ANN inference and training accuracy. In the wake of these adjustments, to perform the ANN MAC operations in the stochastic format, *ODIN* employs the following hardware modifications:

(i) *Lightweight add-on logic circuits* are included in every PCRAM bank to enable the conversion of operands between the binary and stochastic formats. The add-on logic circuits include an SRAM based lookup table for binary to stochastic conversion, and a pop counter for stochastic to binary conversion, of ANN MAC operands. Fig. 4(b)-4(c) shows these add-on logic circuits as part of the schematic of a PCRAM bank. The lookup table consists of an SRAM block of 256x256 cells (Fig. 4(c)), with peripherals to access the block. An 8-bit operand in the binary format is used to index into the SRAM block of 256 rows via a decoder. The indexed SRAM row provides the stochastic version of the operand, which is read in the SRAM block's local row buffer. Along the same lines, the pop counter logic includes a clocked 256-bit parallel-in-serial-out (PISO) register connected with an 8-bit digital level counter (Fig. 4(b)). The PISO stores a 256-bit stochastic operand and then serially outputs it bit-by-bit, which is fed in the counter to count the number of 1s in the stochastic operand, to consequently convert the operand in the binary format. From Fig. 4(a), *ODIN* also employs lightweight control logic to facilitate movement and processing of the operands through these add-on logic circuits. The overheads of these add-on logic circuits and their control logic are discussed in Section V (Table 3).

(ii) *A low-overhead modifications in PCRAM bank peripherals* are included to enable in-situ MAC operations (i.e., series of bit-parallel AND and OR operations) among the operands available in PCRAM in the stochastic format. These modifications are made in the sense-amplifiers, row-address decoders, and wordline drivers, as adopted from PINATUBO framework [3]. Like PINATUBO framework [3], these modifications also enable *ODIN* to perform bit-parallel logical AND/OR/NOT operations between two operands, by simply storing the two operands in two separate PCRAM rows and then activating and reading both the PCRAM rows simultaneously using appropriate sense-amplifier reference voltages. We extract the overheads of these modifications from [3], and account for them in our system-level analysis in Section VI.

*2) Hardware for Activation and Pooling Functions*

To implement binary arithmetic based ANN activation and pooling functions, *ODIN* employs two different add-on logic blocks per physical PCRAM bank (Fig. 4(a)). The logic block for activation function consists of a CMOS implementation of an 8-bit ReLU [24] and it directly follows the pop counter circuit (Fig. 4(a)). On the other hand, the logic block for pooling function implements CMOS circuit for 4:1 8-bit max pooling [25]. The additional control logic included in the PCRAM bank facilitates the movement and processing of operands through these activation and pooling logic blocks. The overheads of activation and pooling logic blocks are given in Table 3 (Section V), and are accounted for in our system level analysis (Section VI). We choose 8-bit ReLU and 4:1 max pooling functions for *ODIN* in this paper, because the ANN benchmark models (Section VI) we have considered for our analysis use these functions. Nevertheless, we envision that *ODIN* can be easily extended to use any other activation (e.g., tanh [26], softmax [27]) and pooling (e.g., average pooling [25]) functions with any other precision as well.

**Table 1: Required number of PCRAM reads, writes, and resultant total latency values for various *ODIN* PIMC commands.**

| *ODIN* PIMC Command | #Reads | #Writes | Latency(ns) |
|---|---|---|---|
| **B_TO_S** | 33 | 32 | 3504 |
| **S_TO_B** | 32 | 32 | 3456 |
| **ANN_POOL** | 32 | 32 | 3456 |
| **ANN_MUL** | 1 | 1 | 108 |
| **ANN_ACC** | 1 | 1 | 108 |

*C. New PIM Controller (PIMC) to Support ANN Processing*

To orchestrate the processing of ANNs by employing the modified hardware discussed in the previous subsection, *ODIN* introduces the following five new PCRAM controller commands: (1) B_TO_S, (2) ANN_MUL, (3) ANN_ACC, (4) S_TO_B, and (5) ANN_POOL. Each of these commands consists of multiple basic PCRAM READ and WRITE operations (Table 1). Fig. 5(a) -5(e) shows PCRAM activity flows undertaken during these commands. Moreover, the latency/timing parameters associated with these commands are given in Table 1. The PIM controller (PIMC) of *ODIN* (Fig. 3) breaks down these PCRAM activity flows into a series of PCRAM operation commands (e.g., READ, WRITE) and control signals. Then, the PCRAM controller schedules these commands in appropriate order while abiding by various timing constraints (e.g., timing constraints of the activity flows in Fig. 5), to orchestrate efficient ANN processing, as discussed below.

(1) **B_TO_S**: this command orchestrates conversion of ANN operands from the binary format to the stochastic format. The activity flow of this command involves reading operands from PCRAM in the binary format, converting them into the stochastic format, and then writing them back into the PCRAM rows of the Compute Partition (Fig. 5(a)). As each 256-bit PCRAM block read from an 8kb PCRAM row can fetch 32 8-bit operands, the B_TO_S activity flow includes conversion of 32 8-bit binary operands into 32 stochastic operands (256-bit each), and writing them back in 32 separate PCRAM rows of the Compute Partition.

(2) **ANN_MUL**: this command directs a multiplication operation (i.e., bit-parallel AND) between two ANN operands that are stored in the stochastic format (256-bit each) in two separate PCRAM rows of the Compute Partition. As discussed earlier, *ODIN* adopts PINATUBO framework [3] for implementing bit-parallel AND operations. As a result, the ANN_MUL activity flow involves simultaneously activating the two PCRAM rows that contain the two 256-bit stochastic operands in the Compute Partition, and then reading both operands simultaneously using an appropriate reference voltage in the sense amplifiers that facilitate reading in the PCRAM bank, as doing so performs bit-parallel AND between the two 256-bit stochastic operands [3]. The reader is encouraged to review [3] for more details on how such bit-parallel AND is performed in the PCRAM sense-amplifiers.

(3) **ANN_ACC**: this command directs an accumulation (addition) operation (i.e., a multiplexing operation converted into a series of bit-parallel AND and OR, Fig. 5(c)) between a stochastic operand stored in a PCRAM row of the Compute Partition and an accumulator block that is part of the Accumulator Row in the Compute Partition. From Fig. 2(b), in addition to the stochastic operands, a multiplexing operation also requires S and S' operands.

*ODIN* pre-processes S and S' operands offline and makes them available in two separate PCRAM rows of the Compute Partition. The ANN_ACC activity flow (Fig. 5(c)) uses these S and S' operands to perform two bit-parallel AND operations and one following bit-parallel OR operation, to consequently perform a multiplexing (accumulate or addition) operation. Like ANN_MUL, ANN_ACC also adopts PINATUBO to undertake its activity flow.

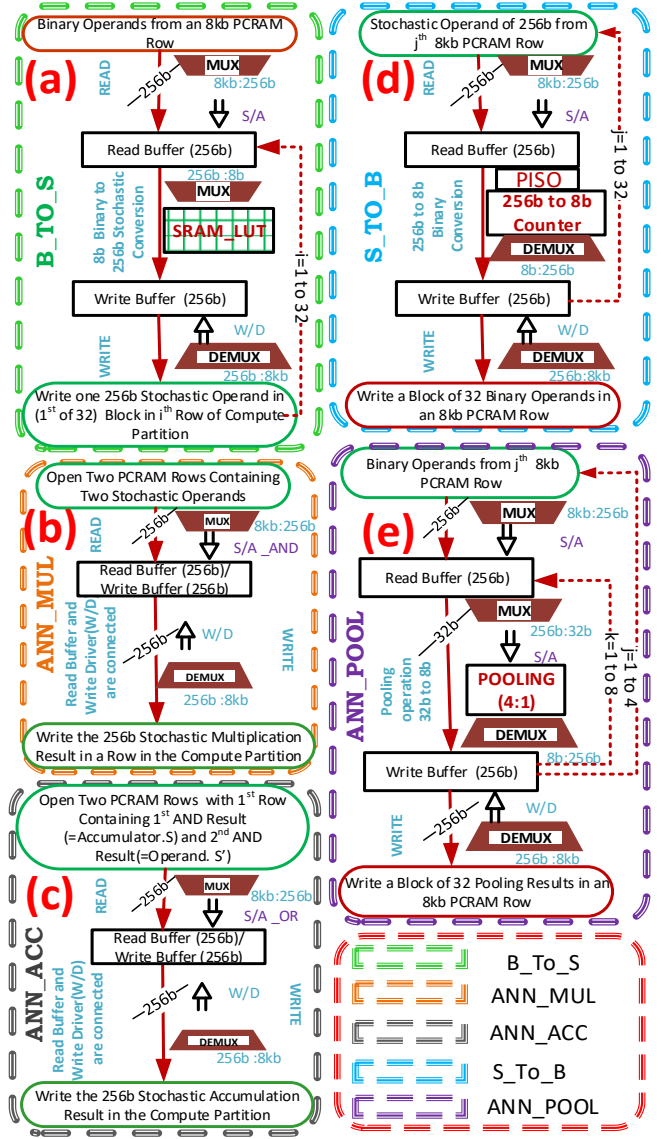

**Fig. 5:** PCRAM activity flows for *ODIN* PIMC commands (a) B_TO_S, (b) ANN_MUL, (c) ANN_MUL, (d) S_TO_B, and (e) ANN_POOL.

(4) **S_TO_B**: After the results of all MAC operations related to at least 32 neurons of an ANN layer are available in the stochastic format, *ODIN* invokes this command to convert the results into the binary format and then apply the activation function on them. The S_TO_B activity flow (Fig. 5(d)) involves reading 32 stochastic MAC results from 32 different PCRAM rows of the Compute Partition, converting them one after another into the binary format using the pop count circuit block, applying the activation function on them one after another using the CMOS logic block for ReLU, and then assembling 32 resultant 8-bit binary activation values in the 256-bit write buffer through demultiplexing, before writing them back (from write buffer) into a PCRAM row in a partition other than the Compute Partition.

(5) **ANN_POOL**: In addition to MAC and activation functions, ANNs may also require (e.g., CNNs) pooling functions. CNNs typically employs one pooling layer (either max or average pooling) after every convolution layer [7]. *ODIN* invokes this command to process a CNN pooling layer. The ANN_POOL activity flow (Fig. 5(e)) changes depending on the size of the pooling filter (e.g., 4:1 vs 9:1). Consequently, the activity flow includes reading multiple (either 4 (Fig. 5(e)) or 9) 256-bit PCRAM blocks containing 32 8-bit binary operands, reducing the number of operands by applying the pooling function using the pooling logic block, and then assembling 32 pooling function outputs in a single 256-bit PCRAM block to be written back. The multiple 256-bit PCRAM blocks that are read at the beginning of this activity flow can come from the same or different PCRAM rows.

## V. IMPLEMENTATION AND HARDWARE OVERHEADS

### A. Implementation of ANN Processing on ODIN

Although *ODIN* can be used for processing ANNs during forward propagation (inference) as well as back propagation (training) phases, in this paper we only discuss and analyze how *ODIN's* performance during the inference phase. To begin with ANN processing, *ODIN* first uploads trained and quantized weights of all layers of the ANN in the PCRAM banks in the 8-bit binary format via the DMA controller (Fig. 3). It also uploads the input activation values to the ANN in the 8-bit binary format. After the uploading of ANN weights and inputs via the DMA controller is over, *ODIN* employs the PIM controller (PIMC) to begin processing the ANN layer by layer, starting from the first layer. To process each layer, *ODIN*'s PIMC breaks down the inference task into a sequence of PCRAM commands and control signals (following the activity flows from Section IV), which are then scheduled by the PCRAM controller to orchestrate the ANN layer processing.

**Table 2: Requirement of memory capacity, number of PCRAM reads and writes for implementing various ANN topologies on *ODIN* accelerator.**

|  | Fully Connected Layers | | | Convolution Layers | | | Accuracy (%) |
|---|---|---|---|---|---|---|---|
|  | Memory (Gb) | Read (x10$^6$) | Write (x10$^6$) | Memory (Gb) | Read (x10$^6$) | Write (x10$^6$) |  |
| VGG1 | 1.93 | 247 | 248 | 0.229 | 58.8 | 30.3 | 89.5 |
| VGG2 | 1.96 | 251 | 252 | 0.234 | 60.01 | 30.9 | 88.51 |
| CNN1 | 0.00095 | 1.22 | 1.226 | 0.0002 | 0.62 | 0.32 | 97.45 |
| CNN2 | 0.00098 | 1.254 | 1.257 | 0.00026 | 0.67 | 0.34 | 97.21 |

**Table 3: Area, energy and delay values for various add-on logic circuits for *ODIN* (scaled for 14nm CMOS).**

| Hardware Component | Energy (pJ) | Delay (ns) | Area (mm$^2$) |
|---|---|---|---|
| SRAM-LUT [28] | 0.297 | 0.316 | 0.402 |
| 16:8 Mux [28] | 4.662 | 0.007 | 0.159 |
| 256:8 Mux [28] | 4.72 | 0.0077 | 0.639 |
| 256:32 Mux [28] | 18.6 | 0.0303 | 0.688 |
| 8:32 Demux [28] | 18.64 | 0.0305 | 0.158 |
| 8:256 Demux [28] | 149.19 | 0.242 | 0.493 |
| 256:1024 Demux [28] | 902.8 | 1.465 | 1.266 |
| ReLU Logic [25] | 185 | 4.3 | 0.02 |
| Pooling Logic [25] | 2140 | 39.3 | 3.06 |

The processing time taken by *ODIN* differs for each layer, depending on the type (e.g., fully-connected, convolution, pooling) and size (i.e., #nodes, #connections) of the layer. This is because the type and size of the layer dictates (i) how many PCRAM reads and writes would be required to process the layer, and (ii) what the required storage capacity would be to store the operands. The type, size, and number of layers in an ANN depends on the architecture of the ANN model in use. Therefore, the storage requirement and the processing performance (i.e., delay, energy) of *ODIN* would differ between different ANN architectures. To provide a peek into these dependencies, we have evaluated the required number of PCRAM reads and writes and storage requirement for some benchmark ANN topologies, as shown in Table 2. Our used methodology for deriving these results is discussed in Section VI-A.

### B. Overheads of Hardware Modifications

The overheads (delay, area, and timing overheads) of different hardware modifications made in PCRAM banks as part of our *ODIN* framework are listed in Table 3. In addition, *ODIN* also needs to add PIMC inside the PCRAM controller module. To support the PIMC, *ODIN* requires five additional control signals corresponding to the controller commands discussed in Section IV. Overall, the overheads of *ODIN* are significantly less than the overheads of other processing in crossbar memory-based accelerators from prior work. This is because, like the crossbar accelerators from prior work, *ODIN* does not need high-area consuming, power-hungry, and throughput-limiting analog-to-digital converters (ADCs) and digital-to-analog converters (DACs). We account for the overheads from Table 3 in our system-level evaluation presented in the next section.

## VI. SYSTEM-LEVEL EVALUATION

### A. Evaluation Setup

For system-level evaluation, we model our *ODIN* accelerator in the following three steps: *(i)* We extract the energy and delay characteristics of PCRAM using PCRAM datasheets [29], and scale them for 14nm using the scaling guidelines given in [30]; *(ii)* We extract the overhead values listed in Table 3 from CACTI [28] based modeling (for Mux, Demux, SRAM) and CMOS custom logic implementations (for ReLU, Pooling) reported in [25] (scaled for 14nm node); *(iii)* We model the execution of ANN benchmark topologies from MLBench (Table 4) using our in-house transaction-level simulator, which incorporates the energy, delay values from steps (i) and (ii).

We compare the performance of *ODIN* with four systems: (i) baseline CPU with 32-bit float precision (referred to as "32-bit CPU"), (ii) CPU with 8-bit fixed precision (referred to as an "8-bit CPU"), (iii) ISAAC with pipeline [2], and (iv) ISAAC without pipeline [2]. We use CPU-only systems as baseline, because prior work [20] also compare their accelerator design with such CPU-only systems. To simulate/evaluate the CPU-only baseline systems, we used gem5 tool [32] with integrated MCPAT (for power modelling).

**Table 4: ANN benchmark topologies [20].**

| CNN1 | conv5x5-pool-784-70-10 (**MNIST Dataset**) |
|---|---|
| CNN2 | conv7x10-pool-1210-120-10 (**MNIST Dataset**) |
| VGG1 | conv3x64-conv3x64-pool-conv3x128 conv3x128-pool-conv3x256-conv3x256-conv3 x 256-pool-conv3 x512-conv3x512-conv3x512-pool-conv3 x512-conv3x512 conv3x512-pool-25088-4096-4096-1000 (**ImageNet Dataset**) |
| VGG2 | conv3x64-conv3x64-pool-conv3x128 conv3x128-pool-conv3x256-conv3x256-conv3x256-conv1 x 512-pool-conv3x512-conv3x512-conv3x512-con1 x512-pool-conv3x512-conv3x512-conv3x512-con1x 512-pool-25088-4096-4096-1000 (**ImageNet Dataset**) |

To simulate/evaluate ISAAC variants, we employ PIMSim [32] with power/energy modeling results for crossbar memories taken from [20]. Moreover, we train our considered ANN benchmark topologies listed in Table 4 with MNIST (CNN-1 and CNN-2) and ImageNet (VGG-1 and VGG-2) datasets. We also quantize the topologies for our 8-bit *ODIN* and other considered accelerator variants. Table 2 lists the inference accuracy of 8-bit quantized topologies.

### B. Evaluation Results and Discussion

Fig. 6(a) and 6(b), respectively, represent the energy consumption and execution time values for CNN-1/2 and VGG-1/2 topologies for our considered baseline and accelerator systems. These results are normalized to *ODIN*. Note that Y-axis on both figures are in log scale. We observe that our proposed *ODIN* accelerator can be up to 5.8× (90.8×) faster and 1554× (23.2×) more energy-efficient compared to the ISAAC variants, across VGG-1/2 (CNN-1/2) topologies. Similarly, *ODIN* can be up to 438× (569×) faster and up to 1530× (30.6×) more energy-efficient compared to the baseline systems, across VGG-1/2 (CNN-1/2) topologies. We also observe that *ODIN* outperforms both CPU-only baseline systems and both ISAAC variants by a large margin for CNN-1/2 topologies. Our *ODIN* accelerator outperforms the other systems for VGG-1/2 topologies as well, but the margin in this case is smaller than in the case of CNN-1/2 topologies. This is because VGG-1/2 topologies require very large number of MAC operations compared to CNN-1/2 topologies, and *ODIN* incurs substantially high delay and energy overheads of converting large number of MAC operands between the binary and stochastic formats. Nevertheless, *ODIN* still outperforms the other considered systems. This is because *ODIN* can access a large amount of data (operands) in situ inside PCRAM-based main memory, as opposed to the other considered systems that have to transfer data (operands) between their main memory and compute units.

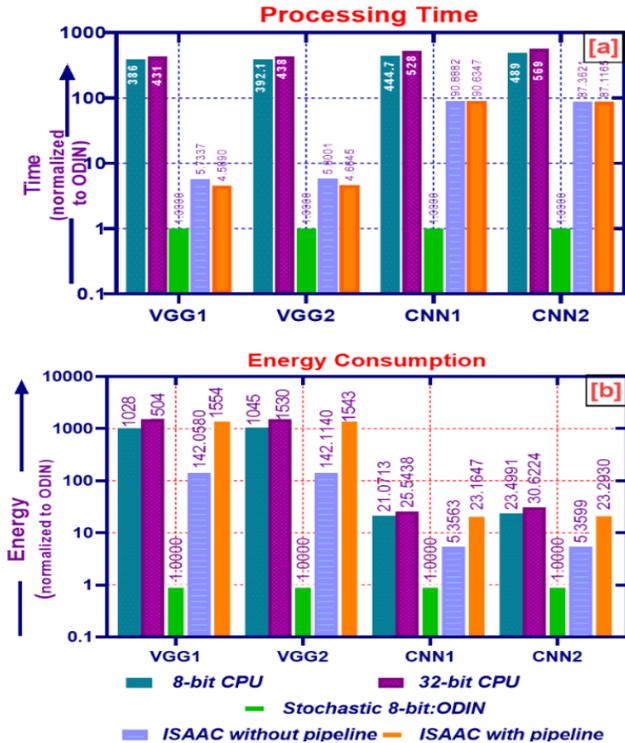

**Fig. 6: (a) Execution time and (b) energy consumption for *ODIN* and other considered ANN processing systems, across VGG-1/2 and CNN-1/2 topologies.**

### CONCLUSIONS

In this work, we presented an energy-efficient and high-throughput ANN accelerator called *ODIN*, which is a PCRAM-based processing-in-memory (PIM) engine. *ODIN* leverages the hybrid binary-stochastic arithmetic to accelerate the essential operations of ANNs, such as multiply-accumulate (MAC), activation, and pooling, directly inside the banks of PCRAM main memory. We analyzed the performance of *ODIN* in terms of execution time and energy consumption, and compared it with two baseline CPU-only systems and two ISAAC-based variants, for four different benchmark ANN topologies from MLBench. The results of our analysis for the considered ANN topologies indicate that our *ODIN* accelerator can be at least 5.8× faster and 23.2× more energy-efficient, and up to 90.8× faster and 1554× more energy-efficient, compared to the crossbar-based in-situ ANN accelerator from prior work. These results corroborate the excellent capabilities of *ODIN* for accelerating ANNs.